\documentclass[aps,twocolumn,
]{revtex4}

\usepackage[dvips]{graphics,epsfig}
\usepackage{amsmath,amsfonts,bm}

\begin{document}

\newcommand\<{\langle}
\renewcommand\>{\rangle}
\renewcommand\d{\partial}
\newcommand\LambdaQCD{\Lambda_{\textrm{QCD}}}
\newcommand\tr{\mathrm{Tr}\,}
\newcommand\+{\dagger}
\newcommand\g{g_5}
\newcommand\gt{\widetilde{g}_5}
\newcommand\epsl{\epsilon_L}
\newcommand\epsr{\epsilon_R}
\renewcommand{\slash}[1]{/\hspace{-7pt}#1}
\newcommand\dslash{/\hspace{-5pt}\partial}
\newcommand{\be}{\begin{equation}}
\newcommand{\ee}{\end{equation}}
\newcommand{\ba}{\begin{eqnarray}}
\newcommand{\ea}{\end{eqnarray}}
\newcommand{\bes}{\begin{subequations}}
\newcommand{\ees}{\end{subequations}}
\newcommand{\ord}{{\cal O}}
\newcommand{\bi}{\begin{itemize}}
\newcommand{\ei}{\end{itemize}}
\newcommand{\gev}{~{\rm GeV}}
\newcommand{\tev}{~{\rm TeV}}
\newcommand{\mev}{~{\rm MeV}}
\newcommand{\psib}{\overline{\psi}}

\title{Extra Gauge Invariance from an Extra Dimension
}
%
\author{Christopher D. Carone}
\author{Joshua Erlich}
\author{Marc Sher}

\affiliation{Particle Theory Group, Department of Physics,
College of William and Mary, Williamsburg, VA 23187-8795}

\date{February 2008}

\newcommand\sect[1]{\emph{#1}---}

\begin{abstract}

We describe higher-dimensional 
theories whose low-energy 4D descriptions contain larger 
gauge or global symmetry groups. As an example, we construct a 
Higgsless SU(2)$\times$U(1)$_{B-L}$ model of electroweak symmetry breaking. 
The 5D SU(2) gauge invariance contains both the weak SU(2) gauge group and a 
custodial symmetry that protects the rho parameter.  Fermions obtain
isospin-violating masses while maintaining universal gauge couplings among 
all three generations. As a further example, we construct a model of chiral
color based on a single SU(3).   
\end{abstract}
%
%
%
\maketitle

{\em Introduction.}--- 
There is by now an established history of extra-dimensional model building, in which 
four-dimensional (4D) effective theories are embedded into higher-dimensional theories.  
Depending on the goals of the model, fields are assumed to propagate 
either through the higher-dimensional bulk, or else along lower-dimensional
branes.  The interest in such models is motivated by phenomenologically
desirable properties of higher-dimensional theories that are absent in 
4D theories, as in 5D models of electroweak symmetry breaking that address the 
hierarchy problem of the standard model~\cite{ADD,RS1}.

In extra-dimensional models, the lighest Kaluza-Klein (KK) modes give 
rise to an effective 4D theory, while the masses of the excited KK 
modes determine the scale above which the extra dimensions become 
manifest. Below the KK scale, a 5D theory with gauge group $G$ 
typically describes a 4D theory with gauge group $G$ or smaller.
However, it is possible that the first few massive modes
are much lighter than the rest, so that the effective 4D description
is that of a larger, product gauge group that is spontaneously broken
to a subgroup of $G$.  Alternatively, the higher-dimensional gauge
invariance may lead to additional global symmetries in the 4D theory.
For example, a 5D SU(2) gauge invariance may contain both the weak SU(2) 
gauge group and an SU(2) custodial symmetry
that protects the $\rho$ parameter.  We describe a framework for realizing 
scenarios of this type, with enhanced gauge or global symmetries in the 
low-energy effective theory.

The basic mechanism that leads to a hierarchy of KK scales is the
existence of multiple ultraviolet (UV) regions in the spacetime geometry.
Models with multiple UV regions and ultralight modes have been studied in
various contexts in the past~\cite{multigravity,KKpar}.  In the case 
of holographic QCD models, the induced geometry on the D8-branes in the 
D4-D8 system~\cite{SS} has two UV regions, as does the deconstructed 
holographic QCD model of Son and Stephanov~\cite{Son-Stephanov}.  
Holographic technicolor models based on the D4-D8 system have been constructed
\cite{SS-tech1,SS-tech2}, and are the most direct analogy
to the models we are considering.

{\em A toy example.}--- 
Consider a 5D SU(2) gauge theory in the background of two slices of
4D anti-de Sitter space joined together in the infrared (IR).  The metric is
\begin{equation}
ds^2=\frac{1}{(1-|z|)^2}\left(\eta_{\mu\nu}\,dx^\mu dx^\nu - dz^2\right), 
\label{eq:metric}\end{equation}
where $\mu,\nu$ run over $0,\dots,3$ and $-1+\epsl<z<1-\epsr$, in
units of the IR length scale.  The geometry has two ultraviolet (UV) 
regions, the neighborhoods of $z_L=-1+\epsl$ and $z_R=1-\epsr$.  In the 
following we will define the conformal 
factor as 
\begin{equation}
w(z)\equiv (1-|z|)^{-1}. 
\end{equation}
The neighborhood of $z=0$, where the conformal factor is smallest, is the IR 
region of the geometry.  We are mainly concerned with the gauge sector of the 
theory, although we acknowledge the potential difficulties in coupling such a model
to gravity.  If coupled to 5D gravity, the
metric (\ref{eq:metric}) is gravitationally unstable due to a ghostlike
radion mode, as was recently stressed in Ref.~\cite{KKpar}.  
The model may instead be coupled directly to 4D gravity, or 
perhaps deconstructed~\cite{deconstruction},
eliminating the instability by fiat; 
it may also be possible to stabilize the geometry by embedding
it in still more dimensions.  

The action for the SU(2) gauge fields is given by
 \begin{equation}
S=-\frac{1}{4}\int d^4x\,dz\,w(z)\,F_{MN}^aF^{a\,MN}, 
\label{eq:Action1}
\end{equation}
where $a$ is the gauge index; the Lorentz indices $M,N$ run 
over 0,1,2,3,$z$ and their contractions are with the flat 5D metric.  
Boundary conditions are chosen to eliminate extra massless scalars
(often referred to as the $A_5$ modes). We will generally choose the
gauge $A_5^a=0$ \cite{higgsless1}.  

The equations of motion for the SU(2) gauge fields $A_\mu^a(x,z)$ are,
\begin{equation}
\partial_z\left[w(z)\partial_z A_\mu^a(x,z)\right]=w(z)\,\d_\nu\d^\nu 
A_\mu^a(x,z). \label{eq:EOM1}\end{equation}
With a common abuse of notation, the mode solutions are of the form
\begin{equation}
A_\mu^a(x,z)=A_\mu^a(x)\,\psi^a(z)\,, 
\end{equation}
where $A_\mu(x)$ satisfies $\partial_\nu\partial^\nu A_\mu^a(x)=-m^2\,A_\mu^a(x)$.

The spectrum depends on the choice of boundary conditions at the 
two UV boundaries.  With Neumann-Neumann boundary conditions $F_{z\mu}^a=0$ ({\em i.e.},
$\partial_z\,A_\mu^a=0$ in the $A_5^a$=0 gauge), there are two light modes.  One is the 
zero mode $m=0$, $\psi_0(z)=$ constant.  The other is an ultralight mode whose mass 
vanishes in the limit $\epsl,\epsr\rightarrow 0$.  The wavefunction of the ultralight
mode is odd about $z=0$ if $\epsl=\epsr$.  The excited KK modes have
masses that, if $\epsl,\epsr\ll 1$, approximate the spectrum of 
the theory with infinite UV cutoff $\epsl,\epsr=0$.  Expanding the solutions
to the equations of motion to leading order in $1/\ln(\epsilon_{L,R})$,
we find that the mass of the ultralight mode is approximately
\begin{equation}
M_Z^2\approx - \frac{\ln(\epsilon_L\epsilon_R)}{\ln\epsilon_L\ln\epsilon_R}\label{eq:zmass}.
\end{equation}
If the boundary conditions on all components of the SU(2) gauge fields were
Neumann-Neumann, then the effective 4D theory would have the spectrum of a 
spontaneously broken SU(2)$\times$SU(2) gauge theory.  With Neumann-Dirichlet boundary 
conditions, the zero mode is lifted, and there is a single ultralight mode with mass 
given by, for small $\epsilon_{L,R}$,
\begin{equation}
M_W^2\approx -(\ln\epsilon_L)^{-1} \,.
\label{eq:wmass}
\end{equation}

{\em A higgsless double-UV model.}--- 
If $A_\mu^3$ satisfies Neumann-Neumann boundary conditions and $A_\mu^{1,2}$ satisfies 
Neumann-Dirichlet boundary conditions, then the toy model described above resembles a 
model of electroweak symmetry breaking, in which the Z boson is the first excited KK mode
of the photon.  However, to obtain the correct standard model
fermion couplings we also gauge a U(1) symmetry.  To couple
fermions correctly without extraneous ultralight modes, we found it easiest
to embed the model in six dimensions.

\begin{figure}
\includegraphics[scale=.8]{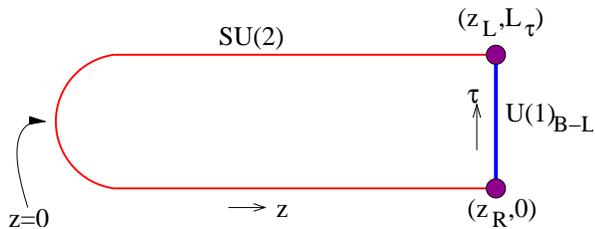}
\caption{\label{fig:one} The D4$_{A}$-D4$_{B}$ brane system.}
\end{figure}

The SU(2) gauge fields propagate along a (five-dimensional) 4-brane D4$_{A}$
with induced metric given by Eq.~(\ref{eq:metric}).  
The U(1) gauge field $B_\mu$ propagates along a 4-brane D4$_B$ which spans 
4D Minkowski space and ends on two points along D4$_A$.  We parametrize the 
extra-dimensional coordinate along D4$_B$ by $\tau$, and we assume that the 
induced metric on D4$_B$ is flat, 
\begin{equation}
ds^2_{D4_B}=\eta^{\mu\nu}dx_\mu dx_\nu-d\tau^2.
\end{equation}
The two intersections of D4$_A$ and D4$_B$ are at $(z,\tau)=
(z_L,L_\tau)$ and $(z_R,0)$ (See Fig.~\ref{fig:one}).  
This scenario is similar to the D4-D8 scenario of Ref.~\cite{SS-tech2}.

We will assume that $L_\tau\lesssim 1$. For the ultralight modes, the U(1) 
gauge field will be approximately uniform in $\tau$, $B_\mu(x,\tau)\approx B_\mu(x)$.
The boundary conditions on the SU(2) gauge bosons $A_\mu^a(x,z)$ and
$B_\mu(x,\tau)$ are determined by beginning with Neumann boundary conditions 
$F_{A\,z\mu}=F_{B\,\tau\mu}=0$ and considering the decoupling limit of a 
Higgs doublet with U(1) charge 1/2, localized at the D4$_A$-D4$_B$ 
intersection $(\tau=0,\,z=z_R)$.  (The SU(2) generators are normalized such 
that ${\rm Tr}\,T^aT^b=\frac{1}{2} \delta^{ab}$.) The large Higgs vacuum expectation 
value (vev) enforces 
\begin{eqnarray}\begin{array}{rcc}
\g A_\mu^3(x,z=z_R)-\gt B_\mu(x,\tau=0)&=&0 \\
A_\mu^{1,2}(x,z=z_R)&=&0 \,.\label{eq:bc1}\end{array}
\end{eqnarray}
We have introduced the 5D SU(2) and U(1) couplings, $\g$ and $\gt$, 
respectively. By the approximate uniformity of $B_\mu(x,\tau)$ for the
ultralight modes, we also have
\begin{equation}
\g A_\mu^3(x,z=z_R)-\gt B_\mu(x,\tau=L_\tau)\approx 0. \label{eq:A3B2}
\end{equation} 
For small enough $\epsl$, $\epsr$ and $L_\tau$, the light spectrum is determined 
approximately by the equations of motion for $A_\mu^a$ alone, with boundary conditions
$\partial_z A_\mu^a=0$ at $z=z_L$, $\partial_z A_\mu^3=0$ at $z=z_L$, and 
$A_\mu^{1,2}=0$ at $z=z_R$.  Boundary conditions on the additional
components are chosen to eliminate any extraneous light degrees of freedom, and
are restricted by self-adjointness of the action \cite{higgsless1}:
\begin{equation}
\begin{array}{rcc}
\partial_z A_5^{1,2}(x,z)|_{z=z_R}&=&0 \\
A_5^a(x,z=z_L)=A_5^3(x,z=z_R)&=&0 \\
B_5(x,\tau=0)=B_5(x,\tau=L_\tau)&=&0 
\end{array}
\end{equation}

We will assume the fermions are localized at the intersections of the 
D4$_A$ and D4$_B$ branes.  The left-handed fermions are at the 
``left boundary'' of the D4$_A$ brane, $z=z_L$, because the W$^\pm$ 
wavefunctions vanish on the right.  The right-handed fermions can {\em a priori}
be localized at either boundary, as SU(2) singlets on the left or SU(2) doublets 
on the right. We will take them to be localized on the right, $z=z_R$.
Then the U(1) is identified with the $B-L$ symmetry of the standard model, with 
quark charge $1/6$ and lepton charge $-1/2$ in our normalization.  

{\em Gauge boson spectrum.}---
The spectrum of gauge bosons is determined by the solutions to 
Eq.~(\ref{eq:EOM1}) subject to boundary conditions.
The lightest modes for $A_\mu^1$ and $A_\mu^2$ are identified with the
W bosons W$^{1,2}$, and have mass approximately given by (\ref{eq:wmass}).  
For $L_\tau\lesssim 1$ the spectrum of light neutral gauge bosons is also
approximately as in the previous section.  The zero mode is identified with 
the photon; the ultralight mode is identified with the Z boson, with mass 
approximately given by Eq.~(\ref{eq:zmass}).  

{\em Standard model fermion couplings.}---  
Consider the couplings of the gauge fields to a left-handed fermion doublet $\psi$ with
U(1) charge $q$.   These arise via the localized kinetic term
$\overline{\Psi} (i\dslash+\g\slash{A}^a T^a+\gt q\,\slash{B})\Psi$ at $(z_L,L_\tau)$.
We replace the fields $A_\mu^a(x,z)$ and $B_\mu(x,\tau)$ with their
mode expansion, including only the massless and ultralight modes:
\begin{eqnarray} \begin{array}{rcl}
A_\mu^{1,2}(x,z)&=&W_\mu^{1,2}(x)\psi_W(z) \\
A_\mu^3(x,z)&=&A_\mu(x)\psi_0+Z_\mu(x)\psi_Z(z) \\
B_\mu(x,\tau)&\approx &\frac{\g}{\gt} A_\mu^3(x,z_R).
\end{array} \end{eqnarray}
The 4D gauge boson couplings are given by
\begin{eqnarray}
{\cal L} &\supset& \overline \Psi \left[\g\, {\bf\slash{W}(x)}\ \psi_W(z_L)
+\g\,\slash{A}(x)\,\psi_0 \,(T^3+q) \right. \nonumber \\
&+& \left. \g \, \slash{Z}(x)(\psi_Z(z_L)\,T^3+\psi_Z(z_R)\,q) \right]\Psi , 
\end{eqnarray}
where we have defined ${\bf W}_\mu=(W_\mu^1 T^1+W_\mu^2 T^2)$.  
Assuming that each mode is normalized so that its 4D kinetic term is 
canonical, we can immediately read off the $W$ and photon couplings:
\begin{equation}
g=\g\,\psi_W(z_L)\, , \,\,\,\,\,\,\,\,\,\, e =\g\,\psi_0 \,.
\label{eq:g}
\end{equation}
Here $\psi_0 \approx (-\ln \epsilon_L \epsilon_R)^{-1/2}$ to leading
logarithmic order, assuming $L_\tau\lesssim 1$ and 
$\g\sim\gt$. The Weinberg angle can be defined in several ways which 
are equivalent at tree level in the standard model. We define, 
\begin{equation}
\sin\theta_W\equiv \frac{e}{g}\simeq
\frac{\psi_0}{\psi_W(z_L)} \approx \left[
\frac{\ln(\epsilon_L)}{\ln(\epsilon_L\epsilon_R)}\right]^{1/2}
\label{eq:sintheta1}
\end{equation}
where the last approximation holds to leading order in  
$\epsilon_{L,R}$. The $Z$ couplings are given by
\begin{eqnarray}
{\cal L}_Z &=&  g_5\, \overline\Psi\,\slash{Z}\,\left[\psi_Z(z_R)\,(T^3+q)
  \right. \nonumber \\ &+&
\left.\left(\psi_Z(z_L)-\psi_Z(z_R)\right)T^3\right]\Psi.
\label{eq:SZ1} 
\end{eqnarray}
which can be compared to the standard model Z coupling to a left-handed
doublet with $B-L$ charge $q$,
\begin{equation}
{\cal L}_Z^{SM}=\frac{g}{\cos\theta_W}\overline\Psi\left(
T^3-\sin^2\theta_W\,(T^3+q)\right)\Psi. \label{eq:SZSM}\end{equation}
Comparing Eqs.~(\ref{eq:SZ1}) and (\ref{eq:SZSM}), we identify,
\begin{equation}
\frac{g}{\cos\theta_W}=
\g\left[\psi_Z(z_L)-\psi_Z(z_R)\right]
\end{equation}
and \begin{equation}
\sin^2\theta_W=\frac{\psi_Z(z_R)}{\psi_Z(z_R)-\psi_Z(z_L)}.
\end{equation}
These definitions of the Weinberg angle and the 4D SU(2) coupling $g$
are not identical to the definitions in Eqs.~(\ref{eq:g}) and 
(\ref{eq:sintheta1}), but are consistent to lowest order 
in $\epsl$ and $\epsr$.  To reproduce the standard model 
value of the Weinberg angle, we must choose 
\begin{equation}
\epsl\approx\epsr^{\tan^2\theta_W}. 
\label{eq:epLepR}\end{equation}

We take the left- and right-handed fermions to be localized on opposite
D-brane intersections.  The right-handed fermions are SU(2) doublets, but
the W boson wave functions vanish at their location. Then the fermion couplings 
agree with the standard model (to lowest order in $\epsilon_{L,R}$).

Using the approximate gauge boson masses (\ref{eq:zmass}) and 
(\ref{eq:wmass}), and $\sin^2\theta_W$ from (\ref{eq:sintheta1}), we
calculate the $\rho$ parameter, 
\begin{equation}
\rho\equiv\frac{m_W^2}{m_Z^2\,\cos^2\theta_W} \approx 
1 +{\cal O}\left(1/\ln(\epsilon_{L,R})\right) .
\end{equation}
Hence, we see explicitly that the $\rho$ parameter is protected in this model
for small $\epsl$, $\epsr$. This is a consequence of the hidden SU(2)$_R$ 
symmetry contained in the 5D SU(2) gauge invariance.  In the standard model, 
$\rho=1$ at tree level due to a similar custodial symmetry in the Higgs 
sector.  

If the cutoffs $\epsl$, $\epsr$ are chosen to fix $\sin^2\theta_W$,
then the oblique parameters S, T and U~\cite{PT,GR,HT}, calculated
numerically following the conventions of Ref.~\cite{higgsless-oblique},
vanish in the limit $\epsl,\epsr\rightarrow0$.  With $\epsr=10^{-16}$ and $L_\tau \ll 1$, 
we find $S=3.9$ and $T=0.04$.  Like the simplest Higgsless models, the $S$ 
parameter is too large, assuming natural values of the cutoff.  Approaches employed
in Higgsless models \cite{higgsless-oblique,higgsless-Sparam}
to reduce $S$ 
might be successfully adapted to double-UV models,
but we do not consider that issue further here.

{\em Fermion masses.}---
\label{sec:fermion-masses}
If the left- and right-handed fermions are localized at opposite
D-brane intersections, then we can obtain fermion masses from nonlocal
operators of the form~\cite{g2}
\begin{equation}
S_{m}= \int d^4x\,\overline{\Psi}_L(x)\, W(x)\,M \Psi_R(x) , 
\label{eq:Sm}
\end{equation}
where $M$ is the mass, or more generally the mass matrix, of the
fermion(s), and $W(x)$ is a product of Wilson lines along the two D4-branes
with the same 4D coordinate $x$, 
\begin{equation}\begin{array}{ccc}
W(x)&=&P\exp\left[i\int_{(x,z_L)}^{(x,z_R)}
dz'^M A_M(z')\right] \\
&&\times \,P
\exp\left[i\int_{(x,0)}^{(x,L_\tau)} 
d\tau'^N B_N(\tau')\right], \end{array} 
\label{eq:fermion-mass}
\end{equation}
and the path-ordered integrals are over paths of constant four-dimensional coordinate
$x$ along the D4$_A$ and D4$_B$ branes.  For locality in the effective 4D
theory, we insist that the Wilson lines only connect fermions with the same
4D coordinate $x$.

The operator in Eq.~(\ref{eq:Sm}) is invariant 
under gauge transformations restricted 
to respect the boundary conditions on the gauge fields. This operator gives mass to the
fermions via the mass matrix $M$. The mass matrix may be isospin-violating because the 
right-handed fermions transform only under an abelian subgroup of the bulk
gauge symmetry. One nice feature about the Wilson-line 
approach to fermion masses is that the couplings to the W and Z
bosons are independent of the fermion masses.

{\em Chiral color from a single SU(3).}---
As another example of enhanced gauge invariance, 
we describe a low-energy theory with a spontaneously
broken SU(3)$_L\times$SU(3)$_R$ gauge symmetry that arises from a 5D SU(3)
gauge theory with the double-UV geometry.   Conventional chiral color
models extend the color group of the standard model to $G = {\rm SU(3)}_L \times {\rm SU(3)}_R$, 
with left- and right-handed quarks transforming in the fundamental representation of the $L$ 
and $R$ gauge groups, respectively~\cite{framp}.  The symmetry $G$ is broken to its diagonal subgroup 
by a Higgs field in the bi-fundamental representation $\Phi\sim({\bf 3},{\bf \overline{3}})$, 
with a vev $\langle \Phi \rangle = (v/\sqrt{2}) \, \openone$. The 
unbroken gauge group is the color group of the standard model. The broken generators correspond 
to a color octet of heavy vector bosons $A$, that are a linear combination of the $G$ gauge 
fields, $A=\cos\theta A_L-\sin\theta A_R$, where $\sin\theta = g_R/\sqrt{g_L^2+g_R^2}$ and 
$m_A = v\sqrt{g_L^2+g_R^2}/\sqrt{2}$. The coupling of the massive,  color-octet fields to 
the left- and right-handed quarks are given by $g_s \cot\theta$  and $-g_s \tan\theta$, 
respectively, with $\tan\theta = g_R/g_L$.  For the case $g_R=g_L$, the $A$ field has purely 
axial-vector couplings to quarks, and is conventionally called an axigluon.  Collider searches 
for new particles decaying to dijets exclude axigluons in the range $200$~GeV$<m_A<980$~GeV~\cite{CDF}.

Now consider a 5D SU(3) gauge theory on an interval bounded by two UV branes, 
at coordinates $z_L$ and $z_R$, with an IR brane located at $z=0$.  We assume Neumann boundary 
conditions on the gauge field at both ends of the interval.  Left- and right-handed
fermions are localized at $z_L$ and $z_R$, respectively. As in our model of electroweak symmetry 
breaking, the zero mode has a flat profile and couples identically to matter at the left and right 
boundaries.  We identify the zero-mode gauge field as the gluon; the strong coupling constant $g_s$ 
is given by
\begin{equation}
g_s = g_5 \frac{1}{(-\ln\epsilon_L\epsilon_R)^{1/2}} \,\,\, ,
\end{equation}
where $g_5$ is the 5D gauge coupling, and the logarithmic factor is fixed entirely by the 
normalization of the zero mode.   On the other hand, the ultra-light massive mode couples to 
matter on the left and right boundaries with the couplings $g_5 \psi_Z(z_L)$ and  $g_5 \psi_Z (z_R)$, 
respectively, in the notation of our electroweak model.  These couplings may be re-expressed in the 
small $\epsilon_{L,R}$ limit as
\begin{equation}
g_L = g_s \left(\frac{\ln\epsilon_R}{\ln\epsilon_L}\right)^{1/2} \,\,\,\,\,
\mbox{ and } \,\,\,\,\,
g_R = - g_s \left(\frac{\ln\epsilon_L}{\ln\epsilon_R}\right)^{1/2} \,\,\, .
\end{equation}
Comparing to the couplings of the massive color octet in the 4D SU(3)$_L \times$SU(3)$_R$ model, one 
identifies
\begin{equation}
\tan\theta = \frac{\ln\epsilon_L}{\ln\epsilon_R} \,\,\,.
\end{equation}
Thus, if $\epsilon_L=\epsilon_R$, the ultra-light mode in this double UV model is indistinguishable
from a conventional axigluon; the effective 4D theory below the mass of the (much heavier) second
KK mode is identical to the 4D SU(3)$_L\times$SU(3)$_R$ model.  The mass of the axigluon in the
extra-dimensional construction is set by choosing the infrared scale.  Quark masses could be obtained
via Wilson line operators or via wave function overlaps in a variant of the model in which quarks
are localized but allowed to propagate into the bulk.

{\em Conclusions.}---
We have shown how 5D theories with double-UV geometries can give 
rise to effective 4D theories with larger gauge or global symmetry 
groups that are broken by boundary conditions.  We showed how to obtain a chiral
color model, usually associated with the gauge group SU(3)$_L\times$SU(3)$_R$
from a 5D theory with a single SU(3) gauge symmetry.  We also showed how a 
model of Higgsless electroweak symmetry breaking could incorporate a custodial 
SU(2) symmetry (which protects the $\rho$ parameter) without requiring that we 
introduce a separate bulk SU(2)$_R$ gauge group.  Additional physics involving product 
group gauge or global symmetries may also be considered in this framework. 
A particularly tantalizing possibility is that the standard model, itself a theory 
with product gauge group, unifies into a higher-dimensional SU(3) grand unified theory.  
Work in this direction is currently in progress.

C.D.C., M.S., and J.E. thank the NSF for support under Grant Nos.~PHY-0456525, PHY-0554854 and 
PHY-0504442, respectively.  J.E. thanks the Jeffress Foundation for support under Grant No.~J-768.

\end{document}